\documentstyle[preprint,aps,epsfig]{revtex}
\tightenlines

\input{amssym.def}
\input{amssym}
\newcommand {\mm}[1]{\ifmmode{#1}\else{\mbox{\(#1\)}}\fi}
\newcommand{\integer}{\mm{{\Bbb Z}}}

\begin{document}
\draft

\title{Statistical Geometry of Packing Defects of Lattice Chain Polymer from Enumeration and Sequential Monte Carlo Method}
\author{Jie Liang$^1$\thanks{Corresponding author. Phone: (312)355--1789, fax: (312)996--5921,
email: {\tt jliang@uic.edu}}, Jinfeng Zhang$^1$ , Rong
Chen$^{1,2}$ }
\address{$^1$Department
      of Bioengineering, SEO, MC-063 \\ and \\
      $^2$ Department of Information and Decision Science\\
      University of Illinois at
      Chicago\\ 851 S.\ Morgan Street, Room 218 \\ Chicago, IL
      60607--7052, U.S.A.}

\date{\today}
\maketitle

\begin{abstract}
Voids exist in proteins as packing defects and are often associated
with protein functions.  We study the statistical geometry of voids in
two-dimensional lattice chain polymers.  We define voids as
topological features and develop a simple algorithm for their
detection.  For short chains, void geometry is examined by enumerating
all conformations.  For long chains, the space of void geometry is
explored using sequential Monte Carlo importance sampling and
resampling techniques.  We characterize the relationship of geometric
properties of voids with chain length, including probability of void
formation, expected number of voids, void size, and wall size of
voids.  We formalize the concept of packing density for lattice
polymers, and further study the relationship between packing density
and compactness, two parameters frequently used to describe protein
packing. We find that both fully extended and maximally compact
polymers have the highest packing density, but polymers with
intermediate compactness have low packing density.  To study the
conformational entropic effects of void formation, we characterize the
conformation reduction factor of void formation and found that there
are strong end-effect.  Voids are more likely to form at the chain
end.  The critical exponent of end-effect is twice as large as that of
self-contacting loop formation when existence of voids is not required.
We also briefly discuss the sequential Monte Carlo sampling and
resampling techniques used in this study.
\end{abstract}
\vspace*{15mm}
\noindent {\bf Keywords:} void, lattice model, lattice packing,
sequential Monte Carlo.

\newpage
\narrowtext

\section{Introduction}

Soluble proteins are well-packed, their packing densities may be as
high as that of crystalline solids
\cite{Richards77_ARBB,Chothia75_Nature,Richards94_QRB}.  Yet there are
numerous packing defects or voids in protein structures, whose size
distributions are broad \cite{LiangDill01_BJ}.  The volume ($v$) and
area ($a$) of protein does not scale as $v \approx a^{3/2}$, which would
be expected for models of tight packing. Rather, $v$ and $a$ scale
linearly with each other \cite{LiangDill01_BJ}.  In addition, the
scaling of protein volume and cluster-radius \cite{Lorenz93_JPA} is
characteristic of random sphere packing.  Such scaling
behavior indicates that the interior of proteins is more like
Swiss cheese with many holes than tightly packed jigsaw puzzles
\cite{LiangDill01_BJ}.

What effects do void have? Proteins are often very tolerant to
mutations
\cite{Lim89_Nature,Shortle90_Biochem,Richards94_QRB,Axe96_PNAS}, which
may suggest potentially stabilizing roles of voids in proteins.  Voids
in proteins are also often associated with protein function.  The
binding sites of proteins for substrate catalysis and ligand
interactions are frequently prominent voids and pockets on protein
structures \cite{Laskowski96_PS,Liang98_PS}.  However, the energetic
and kinetic effects of maintaining specific voids in proteins are not
well-understood, and the shape space of voids of folded and
unfolded proteins are largely unknown.

In this paper, we examine the details of the statistical nature of
voids in  simple lattice polymers.  Lattice models have been
widely used for studying protein folding, where the conformational
space of simplified polymers can be examined in detail
\cite{LauDill89_M,Chan89_M,Dill90,Shak90_JCP,Camacho93_PNAS,Pande94_JPA,Socci94_JCP,Dill95_PS}.
Despite its simplistic nature, lattice model has provided important
insights about proteins, including collapse and folding transitions
\cite{Sali94_Nature,Socci94_JCP,Shriva95_PNAS,Klimov96_PRL,Melin99_JCP},
influence of packing on secondary structure formation
\cite{Chan89_M,ChanDill90_JCP}, and designability of lattice
structures \cite{Gov95_B,Li96_Science}.  However, one drawback is that
lattice model is not well-suited for studying void-related structural
features, such as protein functional sites, since it is not easy to
model the geometry of voids.

In this article, we first define voids as topological defects and
describe a simple algorithm for void detection in two-dimensional
lattice.  We then enumerate exhaustively the conformations for all
$n$-polymers up to $n=22$, and analyze the relationship of probability
of void formation, expected packing density and compactness, as well
as expected wall interval of void with chain length.  To study
statistical geometry of long chain polymers, we describe a Monte Carlo
sampling strategy under the framework of Sequential Importance
Sampling, and introduce the technique of resampling.  The results of
simulation of long chain polymers up to $N=200$ for several geometric
parameters are then presented.  We further explore the conformation
reduction factor $R$ of void formation, and describe the significant
end-effect of void formation, as well as the scaling law of $R$ and
wall interval of voids.  In the final section, we summarize our
results and discuss effective sampling strategy for studying the
conformational space of voids.

\section{Lattice Model and Voids}

Lattice polymers are self-avoiding walks (SAWs), which can be obtained
from a chain-growth model
\cite{Rosenbluth55_JCP,FrenkelSmit96_AP,LandauBinder00}.
Specifically, an $n$-polymer $P$ on a two-dimensional square lattice
$\integer^2$ is formed by monomers $n_i, i \in \{1,..,N\}$.  The
location $x_i$ of a monomer $n_i$ is defined by its coordinates $x_i
= (a_i, b_i)$, where $a_i$ and $b_i$ are integers. The monomers are
connected as a chain, and the distance between bonded monomers $x_i$
and $x_{i+1}$ is $1$.  The chain is self-avoiding: $x_i \ne x_j$ for
all $i \ne j$.  We consider the beginning and the end of a polymer to
be distinct.  Only conformations that are not related by translation,
rotation, and reflection are considered to be distinct.  This is
achieved by following the rule that a chain is always grown from the
origin, the first step is always to the right, and the chain always
goes up at the first time it deviates from the $x$-axis.  For a chain
polymer, two non-bonded monomers $n_i$ and $n_j$ are in {\it
topological contact\/} if they intersect at an edge that they share.
If two monomers share a vertex of a square but not an edge, these two
monomers are defined as not in contact.

\begin{figure}
\vspace{2cm}
  \centerline{\epsfig{figure=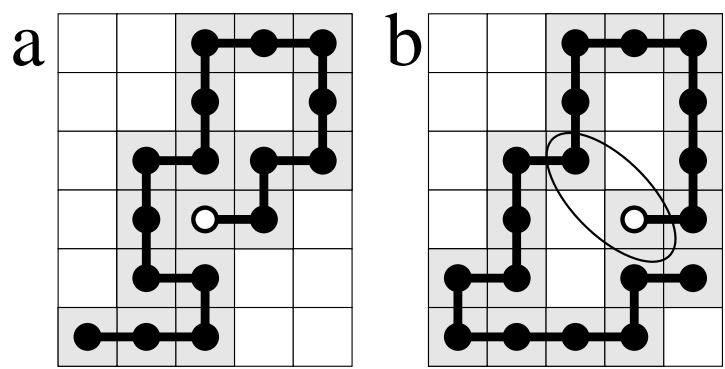,width=4in}}
\vspace{4cm}
\caption{\sf \footnotesize Voids of polymers in square
lattice. Unfilled circle represents the first monomer. (a) A void of
size $1$ is formed in this $17$-mer. (b) The two monomers encircled
shares a vertex but not an edge of a square and are not in topological
contact.  The unfilled space contained within the polymer is regarded
as one connected void of size $4$.}
\label{VoidDef}
\end{figure}

{\bf (Figure~\ref{VoidDef} here.)}

When the number of monomer $n$ is 8 or more, a polymer may contain one
or more void (Figure~\ref{VoidDef}a).  We define voids as topological
features of the polymer.  The complement space $\integer^2 - P$ that
is not occupied by the polymer $P$ can be partitioned into disjoint
components:
\[
\integer^2 - P = \; V_0\; \dot{\cup}\; V_1 \;... \;\dot{\cup}\; V_k.
\]
Here $V_0$ is the unique component of the complement space that
extends to infinity. We call this the {\it outside}.  The rest of the
components that are disjoint or disconnected to each other are {\it
voids\/} of the polymer.  Because non-bonded monomers intersecting at
a vertex are defined as not in contact, they do not break up the
complement space.  As an example, the unfilled space contained within
the polymer in Figure~\ref{VoidDef}b is regarded as one connected void
of size $4$ rather than two disjoint voids of size $2$.  A simple
algorithm for void detection can be found in Appendix.
Figure~\ref{22merVoids4} shows the only six conformations among all
301,100,754 conformations of 22-mer found to have 4 voids.

{\bf (Figure~\ref{22merVoids4} here.)}

\section{Voids Distribution by Exact Enumeration}

{\it Probability of Forming Voids and Expected Number of Voids. \ }  The
number of conformations $\omega(n)$ for $n$-polymer up to $n=25$ by
exhaustive enumeration is shown in Table~\ref{exhaust.tab}.  The
numbers of conformations for polymers up to $n=15$ are in exact
agreement with those reported in Chan and Dill \cite{Chan89_M}.
Table~\ref{exhaust.tab} also lists the number of conformations
$\omega_k(n)$ containing $k= 1, 2, 3$, or $4$ voids.

\begin{figure}[t]
  \centerline{\epsfig{figure=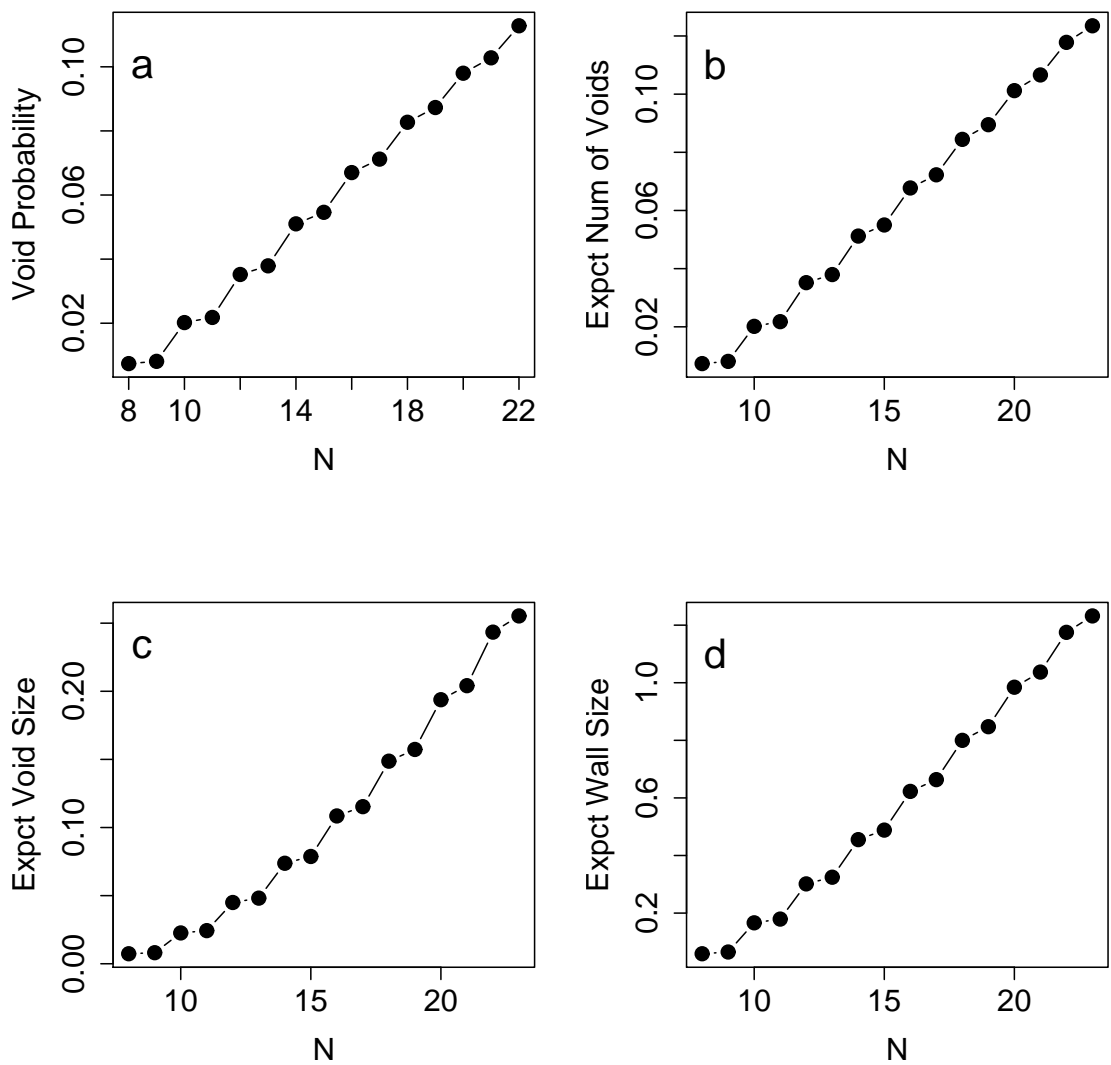,width=6in}}
\vspace{4cm}
\caption{\sf \footnotesize Geometric properties of chain polymers by
exhaustive enumeration.
(a) The probability of void formation,
(b) the expected number of voids contained in a polymer,
(c) the expected void size,
and
(d) the expected wall size of voids. All these parameters
increase with chain length.
}
\label{enum.property}
\end{figure}

The probability for a polymer to form one or more voids $\pi_v$ is
calculated as:
\[
\pi_v =  \frac{\sum_{i=1}^k \omega_k(n)}{\omega(n)}.
\]
The expected number of voids $\bar{n}_v$ for a polymer is:
\[
\bar{n}_v =  \frac{\sum_{i=1}^k \omega_k(n) \cdot k}{\omega(n)}.
\]
As the chain length grows, it is clear that
both $\pi_v$ and $\bar{n}_v$ increases
(Figure~\ref{enum.property}a and Figure~\ref{enum.property}b).

{\bf (Figure~\ref{enum.property} here.)}

{\it Void Size. \ }
The total size $v$ of voids in a polymer is the sum of the  sizes of all
voids, namely, the total number of all unoccupied squares that are
fully contained within the polymer.  Let $\omega_v(n)$ be the number of
conformations of $n$-polymer with total void size $v$.
The expected total void size $\bar{v}$ for $n$-polymer is:
\[
\bar{v} =  \frac{\sum_v \omega_v(n) \cdot v}{\omega(n)}.
\]
Figure~\ref{enum.property}c shows that the expected void size $\bar{v}$
increases with chain length $n$.

{\it Wall Size of Void. \ }
For a void $V$ of size $v$, what is the required minimum length $l(v)$
for a polymer that can form such a void? Equivalently, what is the
size of the wall of the polymer containing void $V$?  Here we first
restrict our discussion to voids formed only by strongly connected
unoccupied sites, namely, any neighboring two sites of a void must be
sharing at least one edge of the squares.  We exclude voids containing
weakly connected sites, where two neighboring sites are connected by
only one shared vertex (Figure~\ref{VoidDef}b).  For $v =1, 2$ and
$3$, it is easy from the geometry of the voids to see that $l(v) = 8,
10$ and $12$, respectively.  However, in general $l(v)$ also depends
on the shape of the void.  A void of size $4$ can have five different
shapes.  If the void is of the shape of a $2 \times 2$ square, $l(4) =
12$.  For the other four shapes, $l(4) = 14$.

For any strongly connected void, we find that the following general
recurrence relationship for $l(v)$ holds:
\[
l(v) = l(v-1) + \left\{
\begin{array}{cc}
2, & {\mbox{if $\Delta \partial V  = 3$}}\\
0, & {\mbox{if $\Delta \partial V = 2$}}\\
-1, & {\mbox{if $\Delta \partial V = 1$}},\\
\end{array}
\right.
\]
where $\partial V$ represents the boundary edges of void $V$, and
$\Delta \partial V$ represents the net gain in the number of
boundary edges introduced by the newly added unoccupied site.
Although the number and explicit shapes of strongly connected voids of
size up to 5 can be found in \cite{Golomb94}, there is no general
analytical formula known for the number of shapes of a void of size
$v$.  This is related to the problem of determining the number of
polyominos or animals (as in percolation theory) of a given size.

\begin{figure}[t]
  \centerline{\epsfig{figure=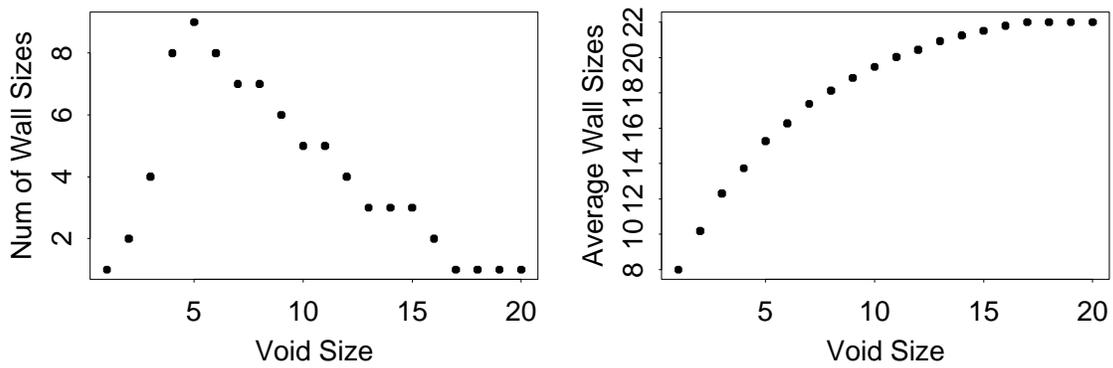,width=6in}}
\vspace{4cm}
\caption{\sf \footnotesize Voids of fixed size in polymers can have
different shapes and thus sometimes different wall sizes.  (a) The
distribution of the number of observed different wall sizes for a void
depends on the size of the void. Voids of size 5 has the maximum number
of different wall sizes. (b) The expected wall size for voids of
different size in 22-mer. }
\label{22merWallhist}
\end{figure}

When weakly connected voids are also considered, there are more
possible wall sizes for void.  For $22$-mer, the number of different
wall sizes observed for a void, strongly or weakly connected, at
various size are shown in Figure~\ref{22merWallhist}a.  Voids of size
5 has the largest diversity in wall size. This is of course due to the
fixed chain length.  A short chain such as the 22mer has only a small
number of ways for form large voids.  Figure~\ref{22merWallhist}b
shows the average wall size for various void size in 22-mer.  The
expected or average wall size $\bar{w}(n)$ for a void in a $n$-polymer
can be calculated as:
\[
\bar{w}(n)
= \sum_{v}w \cdot \frac{  \omega_{v, w}(n)}{\omega_v(n)},
\]
where $v$ is the void size, $w$ the wall size of the void, $\omega_{v,
w}(n)$ is the number of $n$-polymers containing a void of size $v$
with wall size $w$, and $\omega_v(n)$ is the total number of
$n$-polymers with a void of size $v$.  Figure~\ref{enum.property}d
shows that $\bar{w}(n)$ increases with chain length.  Wall size and void
size are analogous to the area and volume of voids in three
dimensional space.

{\it Packing Density. \ }
An important parameter that describes how effectively atoms fill space
is the packing density $p$.  In proteins, it is defined by Richards
and colleagues as the amount of the space that is occupied within the
van der Waals envelope of the molecule, divided by the total volume of
space that contains the molecule
\cite{Richards94_QRB,Richards74_JMB}.
It has been widely used by
protein chemists as a parameter for characterizing protein folding
\cite{Richards94_QRB}.  Following this original definition,
the packing density $p$ for lattice polymer is:
\[
p = n/(n+v),
\]
when a $n$-polymer has a total void size of $v$.

\begin{figure}[t]
  \centerline{\epsfig{figure=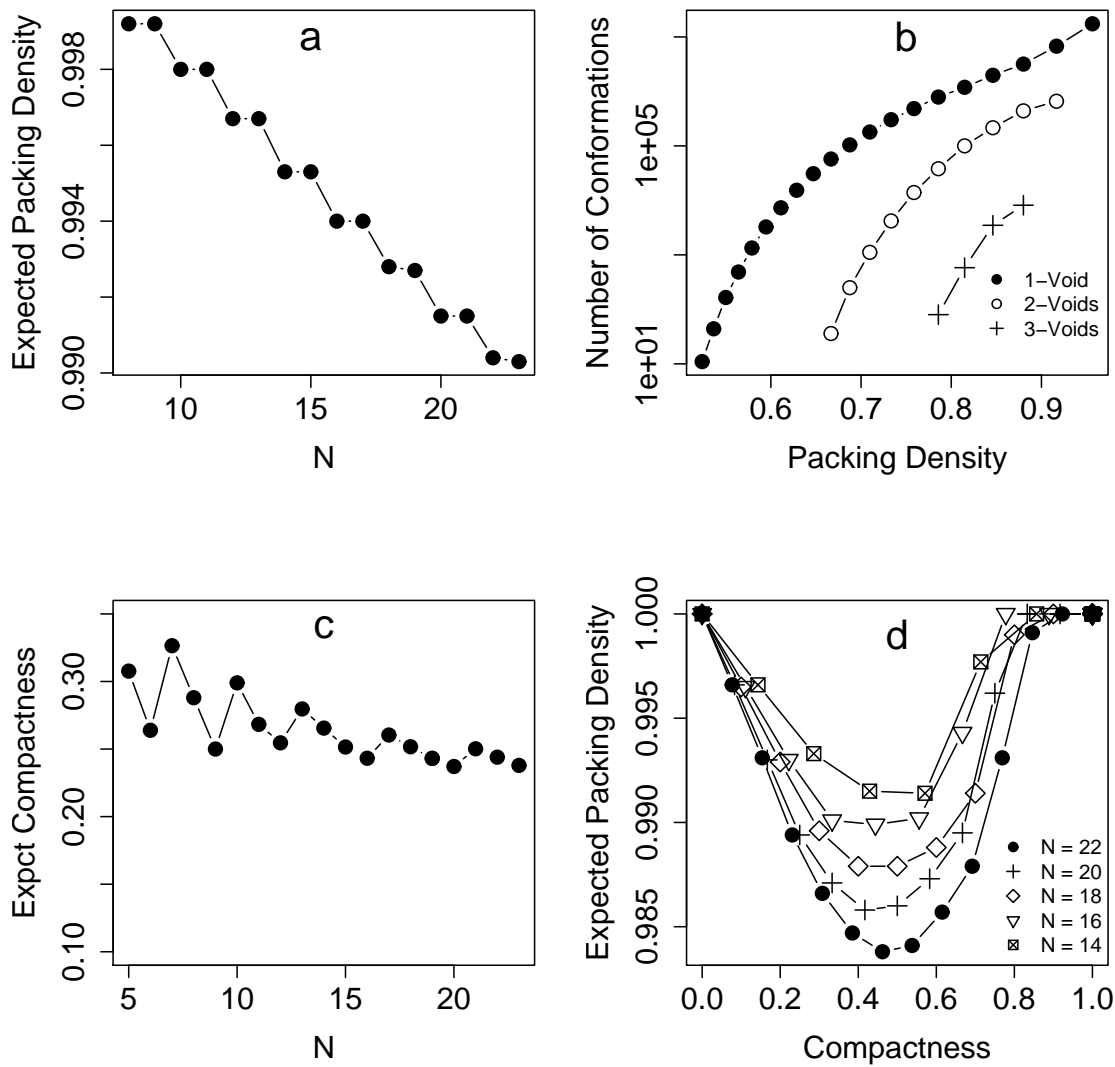,width=6in}}
\vspace{1cm}
\caption{\sf \footnotesize Packing density and compactness are two
useful parameters describing packing of chain polymers.  (a) The
expected packing density decreases with chain length;
(b) For 22-mer, the majority of the conformations
with 1-void have high packing density, namely, the size of void is
small.  Fewer conformations are found with large voids.  The same
pattern is observed for conformations with 2 and 3 voids;
(c) The
expected compactness fluctuates but in general decreases with chain
length;
(d) The
relationship of average packing density $p$ and average compactness
$\rho$ for chain polymer of length $14-22$.  Both maximally compact
polymer ($\rho=1$) and extended polymer ($\rho=0$) have maximal
packing density ($p = 1$), but polymers with low packing density
have intermediate compactness on average.}
\label{pd.compactness}
\end{figure}

The expected packing density $\bar{p}(n)$ for a $n$-polymer can be
calculated as:
\[
\bar{p}(n) = \sum_p p \cdot \frac{ \omega_p(n)}{\omega (n)},
\]
where $\omega(n)$ is the number of all conformations of $n$-mer,
$\omega_p(n)$ the number of $n$-mers with packing density of $p$.  The
scaling of $\bar{p}(n)$ with the chain length $n$ decreases roughly
linearly between $n=7$ and $n=22$ (Figure~\ref{pd.compactness}a).
Because it takes at least two additional monomer to increase the size
of a void by one, $\bar{p}(n)$ decreases only when $n$ is an odd
number for short chains.

Although voids are packing defects, most conformations with voids have
high packing density, namely, the total size of voids are small. Among
all conformations of 22-mer containing one void, the number of
conformation increases monotonically with packing density.  The lowest
packing density $0.52$ has only $11$ conformations, whereas the
highest packing density $0.92$ has the largest number ($6,756,751$) of
conformations (Figure~\ref{pd.compactness}c).  Similar relationship is
found among conformations with 2 and 3 voids
(Figure\ref{pd.compactness}c).

{\it Compactness. \ }  Another important parameter that measures packing
of lattice polymer is the number of nonbonded contacts $t$.  It is
related to the {\it compactness\/} parameter $\rho$, defined in
\cite{Chan89_M} as $\rho = t/t_{\max}$, where $t_{\max}$ is the
maximum number of nonbonded contact possible for a $n$-polymer.
Compactness $\rho$ has been studied extensively in seminal works by
Chan and Dill \cite{Chan89_M,ChanDill90_JCP,ChanDill89_JCP}.  Although
$p$ is sometimes correlated with the compactness $\rho$, these two
parameters are distinct.  The relationship between compactness and
expected packing density for chain polymer of length $14-22$ is shown
in Figure~\ref{pd.compactness}.  For all chain lengths, both maximally
compact polymer ($\rho=1$) and extended polymer ($\rho=0$) have
maximal packing density ($p = 1$). Polymers with $\rho$ between 0.4
and 0.6 have lowest packing density and therefore tend to have larger
void size.  The explanation is simple. An extended lattice chain
polymer has no voids, it therefore achieves maximal packing density of
$p = 1$, but its compactness $\rho$ is 0.  A maximally compact polymer
with $\rho=1$ also contains no voids, its $p$ is $1$.  On the other
hand, non-maximally compact polymers can have a range of packing
densities.

{\bf (Figure~\ref{pd.compactness} here.)}

\section{Obtaining Void Statistics for Long Chains via Importance Sampling}

{\it Sequential Importance Sampling.}  Geometrically complex and
interesting features emerge only in polymers of sufficient length,
which are not accessible for analysis by exhaustive enumeration, due
to the fact that the number of possible SAWs increases exponentially
with the chain length. Monte Carlo methods are often used to generate
samples from all possible conformations and obtain estimates of
feature statistics using those samples. However, when chain length
becomes large, the direct generation of SAWs using rejection method
({\it i.e.}, generate random walks on the lattice and only accept
those that are self-avoiding) from the uniform distribution of all
possible SAW's becomes difficult. The success rate $s_N$ of generating
SAWs decreases exponentially: $s_N \approx Z_N/(4\times 3^{N-1})$.
For $N=48$, $s_N$ is only $0.79\%$ \cite{Liu01_MC}.  To overcome this
attrition problem, a widely used approach is the Rosenbluth Monte
Carlo method of biased sampling \cite{Rosenbluth55_JCP}. The task is
to grow one more monomer for a $t$-polymer chain that has been
successfully grown from 1 monomer after $t-1$ successive steps without
self-crossing, until $t=n$, the targeted chain length. In this method,
the placement of the $(t+1)$-th monomer is determined by the current
conformation of the polymer.  If there are $n_t$ unoccupied neighbors
for the $t$-th monomer, we then randomly (with equal probability) set
the $(t+1)$-th monomer to any one of the $n_t$ sites. However, the
resulting sample is biased toward more compact conformations and does
not follow the uniform distribution. Hence each sample is assigned a
``weight'' to adjust for the bias.  Any statistics can then be
obtained from weighted average of the samples.  In the case of
Rosenbluth chain growth method, the weight is computed recursively as
$w_t=n_tw_{t-1}$.

Liu and Chen \cite{Liu&Chen98} provided a general framework of
Sequential Monte Carlo (SMC) methods which extend the Rosenbluth
method to more general setting. Sophisticated but more flexible and
effective algorithms can be developed under this framework. In the
context of growing polymer, SMC can be formulated as follows.  Let
$(x_1,\ldots,x_t)$ be the position of the $t$ monomers in a chain of
length $t$. Let $\pi_1(x_1), \pi_2(x_1,x_2),\ldots,
\pi_t(x_1,\ldots,x_t)$ be a sequence of {\it target} distributions,
with $\pi(x_1,\ldots,x_n) = \pi_n(x_1,\ldots,x_n)$ being the final
objective distribution from which we wish to draw inference from. Let
$g_{t+1}(x_{t+1}\mid x_1,\ldots, x_t)$ be a sequence of {\it trial
distributions} which dictates the growing of the polymer.  Then we
have:

\begin{tabbing}
123\=456\=789\=\kill
{\bf Procedure} {\sc SMC} ($n$)\\
Draw $x^{(j)}_1$, $j=1, \ldots, m$ from $g_1(x_1)$\\
Set the incremental weight $w^{(j)}_1 = \pi_1(x^{(j)}_1)/g_1(x^{(j)}_1)$\\
{\tt for } $t =1$ {\tt to} $n-1$ \\
   \> {\tt for } $j =1$ {\tt to} $m$ \\
   \> \> {\sf // Sampling for the $(t+1)$-th monomer for the $j$-th sample}\\
   \> \> Draw position $x^{(j)}_{t+1}$ from \\
   \> \>  \> $g_{t+1}(x_{t+1}|x^{(j)}_1 \ldots x^{(j)}_{t})$\\
   \> \> {\sf // Compute the incremental weight.}\\
   \> \>   $u^{(j)}_{t+1} \leftarrow
\frac{\textstyle{\pi_{t+1}(x^{(j)}_1 \ldots x^{(j)}_{t+1})}}
          {\textstyle{\pi_t(x^{(j)}_1 \ldots x^{(j)}_t)
\cdot g_{t+1}(x^{(j)}_{t+1}|x^{(j)}_1 \ldots x^{(j)}_{t})}}
$
       \\
   \> \>  $w^{(j)}_{t+1} \leftarrow u^{(j)}_{t+1} \cdot w^{(j)}_t$\\
   \> {\tt endfor} \\
   \>{\sf Resampling} \\
{\tt endfor}
\end{tabbing}

At the end, the configurations of successfully
generated polymers $\{(x_1^{(j)},\ldots,x_n^{(j)})\}_{j=1}^m$
and their associated weights $\{w_n^{(j)}\}_{j=1}^m$
can be used to
estimate any properties of the polymers, such as expected void
size, compactness, and packing density. That is, the objective inference
$\mu_h=E_\pi[h(x_1,\ldots,x_n)]$ is estimated with
\begin{equation}
\hat{\mu}_h=\frac{\sum_{j=1}^m h(x^{(j)}_1,\ldots,x^{(j)}_n) \cdot w_n^{(j)}}
{\sum_{j=1}^m w_n^{(j)}},  \label{est}
\end{equation}
for any integrable function $h$ of interests.

The critical choices that affect the effectiveness of the SMC method
are: (1) the approximating target distribution $\pi_{t}(x_1 \ldots
x_{t})$, (2) the sampling distribution $g_{t+1}(x_{t+1}|x_1 \ldots
x_{t})$, and (3) the resampling scheme.  In this study, we are
interested in sampling from the uniform distribution $\pi_{n}(x_1
\ldots x_n)$ of all geometrically feasible conformations of length
$n$, which we call the final objective distribution. It can also be
chosen to be the Boltzmann distribution when energy function such as
the HP model \cite{Dill85_Biochem,LauDill89_M,ChanDill93_JCP} is
introduced.

The Rosenbluth method \cite{Rosenbluth55_JCP} is a special case of
SMC. Its target distributions $\pi_{t}(x_1 \ldots x_{t})$ is the
uniform distribution of all SAWs of length $t$. Its sampling
distribution $g_{t+1}(x_{t+1}|x_1 \ldots x_t)$ is the uniform
distribution among all $n_1(x_1,\ldots,x_t)$ unoccupied neighboring
sites of the last monomer $x_t$, and the weight function is
\[
w(x_1, \ldots, x_t,x_{t+1})=w(x_1,\ldots,x_t)n_1(x_1,\ldots,x_t).
\]
When there is no unoccupied neighboring sites
($n_1(x_1,\ldots,x_t)=0$), there is no place to place the $(t+1)$-th
monomer. In this case, the chain runs into a dead end and we declare
the conformation {\it dead}, with weight assigned to be $0$.  In the
case of Rosenbluth method, no resampling is used.

Similarly, the $k$-step look ahead algorithm
\cite{Meirovitch82_JPA,Liu01_MC} chooses $\pi_{t+1}(x_1,\ldots,
x_{t+1})$ being the marginal distribution of
$\pi^*_{t+k}(x_1,\ldots,x_{t+k})$, the uniform distribution of all
SAW's of length $t+k$. Hence $\pi_{t+1}$ is closer to the final
objective distribution -- the uniform distribution of all SAW's of
length $n$.  Specifically,
\begin{eqnarray*}
\pi_{t+1}(x_1,\ldots, x_{t+1}) & = & \sum_{x_{t+2},\ldots,x_{t+k}}
\pi^*_{t+k}(x_1,\ldots, x_{t+1},x_{t+2},\ldots,x_{t+k}) \\
& \propto & n_k(x_1,\ldots,x_{t+1})
\end{eqnarray*}
where
$n_k(x_1, \ldots, x_{t+1})$ is the total number of SAWs of length $t+k$
``grown'' from $(x_1,\ldots, x_{t+1})$ [i.e.
with the first $(t+1)$ positions at $(x_1,\ldots,x_{t+1})$.]
In the $k$-step look-ahead algorithm, the sampling distribution  is
\[
g_{t+1}(x_{t+1}=x\mid x_1,\ldots, x_t)=
\frac{n_{k}(x_1,\ldots,x_t,x)}{n_{k+1}(x_1,\ldots,x_t)}.
\]
It chooses the next position according to what will happen $k$ steps
later.  Namely, the probability of placing the $t+1$-th monomer at $x$
is determined by the ratio of the total number of SAWs of length $t+k$
grown from $(x_1,\ldots,x_t,x)$ and the total number of SAWs of the
same length $t+k$ grown from one step earlier $(x_1,\ldots,x_t)$.  The
corresponding weight function is
\begin{eqnarray*}
w(x_1,\ldots,x_t,x_{t+1}) &=& 
\frac{n_k(x_1,\ldots,x_{t+1})}{n_{k}(x_1,\ldots,x_{t})\cdot
\frac{n_k(x_1,\ldots,x_{t+1})}{n_{k+1}(x_1,\ldots,x_{t})}
}
\\
& = & \frac{n_{k+1}(x_1,\ldots,x_t)}{n_{k}(x_1,\ldots,x_{t})}.
\end{eqnarray*}
Although it has higher computational cost, it usually produces better
inference on the final objective distribution, with less ``dead''
conformations.  The standard Rosenbluth algorithm is $1$ step look ahead
algorithm.

\begin{figure}[t]
  \centerline{\epsfig{figure=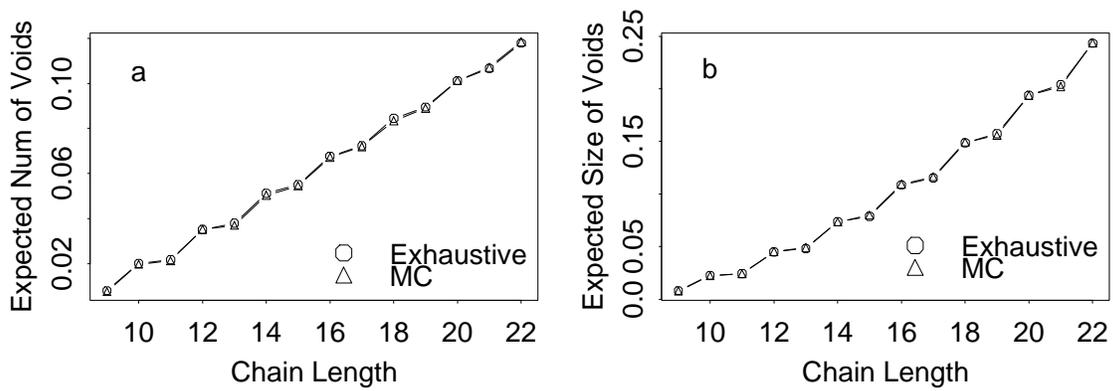,width=6in}}
\vspace{4cm}
\caption{\sf \footnotesize Geometric properties obtained by
enumeration and by Monte Carlo sampling for polymers of chain length
$9-22$.  (a) The expected number of voids, and (b) the expected size
of voids.  Two-step look-ahead sequential Monte Carlo sampling is used,
and the sample size is 100,000.  These data show that geometric
properties estimated by Monte Carlo are identical to those
obtained by exhaustive enumeration.}
\label{exhaustMC.ps}
\end{figure}

To compare geometric properties estimated from sequential Monte Carlo
method and those obtained by exhaust enumeration, we examine the
expected number of voids and expected void size for polymer from
chain length $14$ to $22$.  Figure~\ref{exhaustMC.ps} shows that
sequential Monte Carlo can provide very accurate estimation of these
geometric properties of voids. Here 2-step look ahead is used, with Monte
Carlo sample size of 100,000 and no resampling.

{\bf (Figure~\ref{exhaustMC.ps} here.)}

The resampling step is one of the key ingredient of the SMC
\cite{Liu&Chen95,Liu&Chen98}. There are many cases where resampling
is beneficial.  First, note that it is unavoidable to have some dead
conformations during the growth. These chains need to be replaced to
maintain sufficient Monte Carlo sample size. Second, the weight of
some chains may become relatively so small that their contribution in
the weighted average (\ref{est}) is negligible. When the variance of
the weights is large, the {\it effective Monte Carlo sample size}
becomes small \cite{Kong&94,Liu&Chen95,Liu&Chen98}. Third,
for a specific function $h$, its value may become too small (even
zero) for some sampled conformations. In all these cases, efficiency
can be gained by replacing those conformations with ``better'' ones.
This procedure is called ``resampling''. There are many different ways
to do resampling. One approach is {\it rejection control}
\cite{Liu&98}, which regenerates the replacement conformations
from scratch. An easier approach is to duplicate the existing
and {\it good} conformations \cite{Liu&Chen98}. Specifically,

\begin{tabbing}
123\=456\=789\=\kill
{\bf Procedure} {\sc Resampling}\\
//{\sf $m$: number of original samples.}\\
//{\sf $\{(x_1^{(j)}, \ldots, x_t^{(j)}), w^{(j)}\}_{j=1}^m$: 
original properly weighted samples} \\
{\tt for} $j=1$ {\tt to} $m$\\
\> Set resampling probability of $j$th conformation $\propto \alpha^{(j)\
}$\\
{\tt endfor}\\
{\tt for} $*j=1$ {\tt to} $m$\\
\> Draw $*j$th sample
 from  original samples $\{(x_1^{(j)}, \ldots, x_t^{(j))}\}_{j=1}^m$\\
 \> \>     with probabilities  $\propto \{\alpha^{(j\
 )}\}^m_{j=1}$\\
 \>   //{\sf Each sample in the newly formed sample is assigned a new 
weight.}\\
 \>  //{\sf $*j$-th chain in new sample is a copy of  $k$-th chain in 
original sample.}  \\
 \> $w^{(*j)} \leftarrow w^{(k)}/\alpha^{(k)}$ \\
 {\tt endfor}\\
 \end{tabbing}

In the resampling step, the $m$ new samples $\{(x_1^{(*j)}, \ldots,
x_t^{(*j)}\}_{j=1}^m$ can be obtained either by residual sampling or
by simple random sampling. In residual sampling, we first obtain the
normalized probability $\tilde{\alpha}^{(j)}=\alpha^{(j)}/ \sum
\alpha^{(j)}$. Then $[m\tilde{\alpha}^{(j)}]$ copies of $j$-th sample
are made deterministically for $j=1,\ldots,m$. For the remaining
$m-\sum [m\tilde{\alpha}^{(j)}]$ samples to be made, we randomly
sample from the original set with probability proportional to
$m\tilde{\alpha}^{(j)}-[m\tilde{\alpha}^{(j)}]$.

The choice of
resampling probability proportional to $\alpha^{(j)}$ is problem
specific. For general function $h$, such as the end-to-end extension
$||x_n-x_1||$, it is common to use $\alpha^{(j)}=w_t^{(j)}$. In this
case, all the samples in the new set have equal weight. When the
function is irregular, a carefully chosen set of $\alpha^{(j)}$ will
increase the efficiency significantly.

The method of pruning and enriching of Grassberger
\cite{Grassberger97_PRE} is a special case of the residual sampling,
with $\alpha^{(j)}=0$ for the $k$ chains with zero weight (dead
conformations), $\alpha^{(j)}=2$ for the top $k$ chains with largest
weights, and $\alpha^{(j)}=1$ for the rest of the chains. Residual
sampling on this set of $\alpha$ is completely deterministic.  The
resulting sample consists of two copies of the top $k$ conformations
(each of them having half of their original weight) and one copy of
the middle $n-2k$ chains with their original weight. The $k$ dead
conformations are removed.

In our study of the relationship between compactness and packing
density, we use a more flexible resampling method.  Our focus is on
the packing density among all conformations with certain range of
compactness. In this case, our object target distribution is the
uniform distribution among all possible SAW's with compactness measure
falling within a certain interval, {\it i.e.}, a truncated
distribution. Although compactness changes slowly as the chain grows,
to grow into a long chain it is possible that the compactness of a
chain evolve and cover a wide range during growth. Hence we choose the
uniform distribution of all possible SAWs' of length $t$ as our target
distribution at $t$, and only select those with the desire compactness
at the end for our estimation of the packing density. In order to have
higher number of usable samples ({\it i.e.,}, to achieve better
acceptance rate) at the end, we encourage growth of chains with
desirable compactness through resampling. Specifically,

\begin{tabbing}
123\=456\=789\=\kill
{\bf Procedure} {\sc Resampling} ($m, d, c_t$)\\
// {\sf $m$: Monte Carlo sample size, $d$: steps of looking-back.}\\
// {\sf $c_t$: targeting compactness.}\\
$k \leftarrow$  number of dead conformations.\\
Divide $m - k$ samples randomly into $k$ groups.\\
{\tt for } group $i =1$ {\tt to} $k$ \\
   \> Find conformations not picked in previous $d$ steps.\\
   \>  \>//{\sf Pick the best conformation $P_j$, for example} \\
   \>  \>$P_j \leftarrow$ polymer with $\min |c - c_t|$\\
   \> Replace one of $k$ dead conformations with $P_j$\\
   \> Assign both copies of $P_j$ half its original weight.\\
{\tt endfor}
\end{tabbing}
Here $d$ is used to maintain higher diversity for resampled
conformations.

\begin{figure}[t]
  \centerline{\epsfig{figure=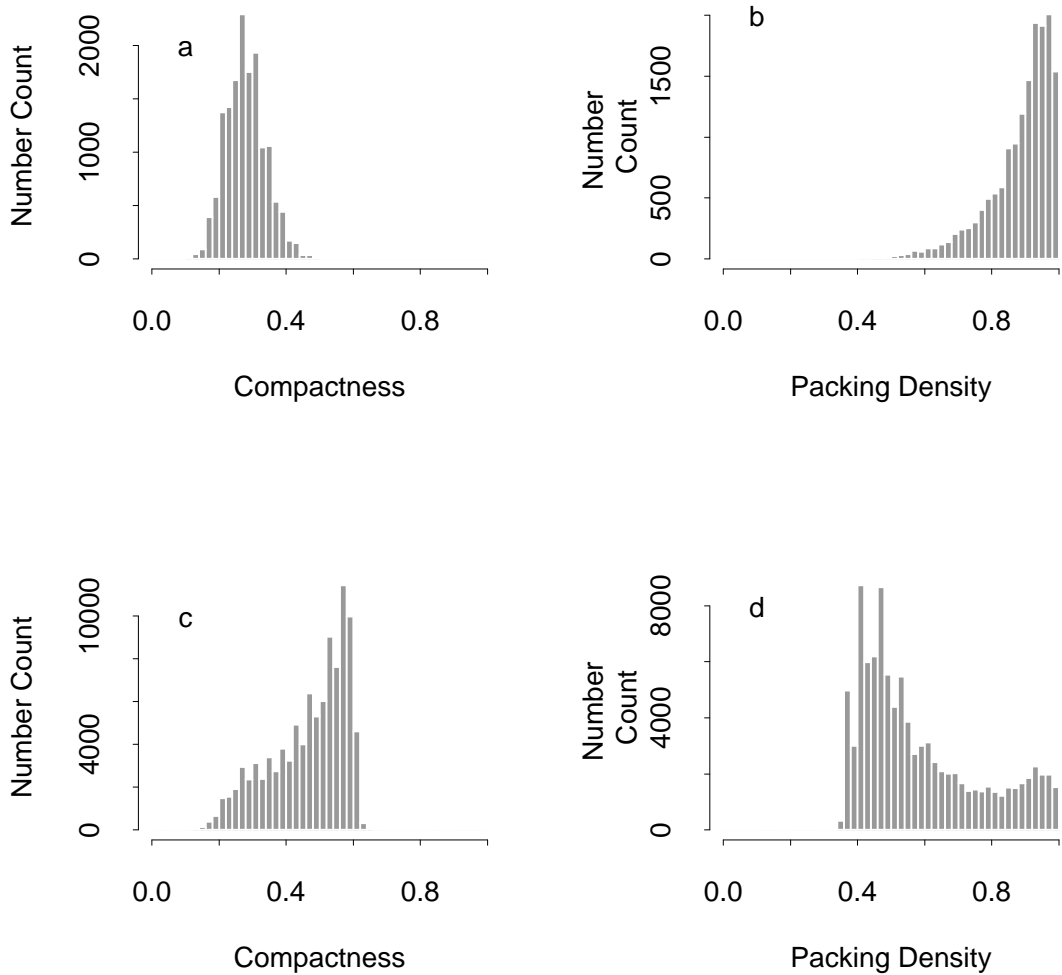,width=6in}}
\vspace{1cm}
\caption{\sf \footnotesize The distribution of configurations of
polymers obtained by sequential Monte Carlo method can be adjusted by
resampling. Sequential Monte Carlo of two step look-ahead without
resampling does not generate enough compact conformations. (a)
Histogram of conformations at different compactnesses generated
without resampling. The compactness of the majority of the
conformations is less than 0.5.  (b) Histogram of conformations at
different packing density generated without resampling. The majority
of the conformations are more extended and have higher packing density.  The
number of conformations with packing density below 0.8 is small. (c)
After applying resampling technique favoring compactness of 0.6, the
majority of the conformations have compactness between 0.5 and 0.6.
Here resampling is applied at each sequential Monte Carlo growth step.
(d) Resampling can also be applied to generate conformations with low
packing densities with voids.  Here resampling favoring low packing
density is applied every 2 growth steps.  Sample size of 200,000 is
used in all calcualtions. 
}
\label{resampling}
\end{figure}

Most polymers sampled by sequential Monte Carlo without resampling are
well-extended with few voids, as shown in Figure~\ref{resampling}(a)
and (b). They have small compactness (less than 0.5) and large packing
density.  As a result, a small number of samples are accepted at the
end whose compactness falls within the desired interval of higher than
0.5.  By using the resampling step described above, we were able to
generate more samples near the desired compactness value of 0.6
(Figure~\ref{resampling}c). Figure ~\ref{resampling}c is a pure
histogram of compactness in the observed samples, without regarding
the weight of the samples. Figure~\ref{resampling}d shows that the
resampling technique is also very effective in shifting the samples to
small packign density values, hence improve the inferences.

{\bf (Figure~\ref{resampling} here.)}

\section{Voids Distribution of Long Chains}
\begin{figure}[!t]
\centerline{\epsfig{figure=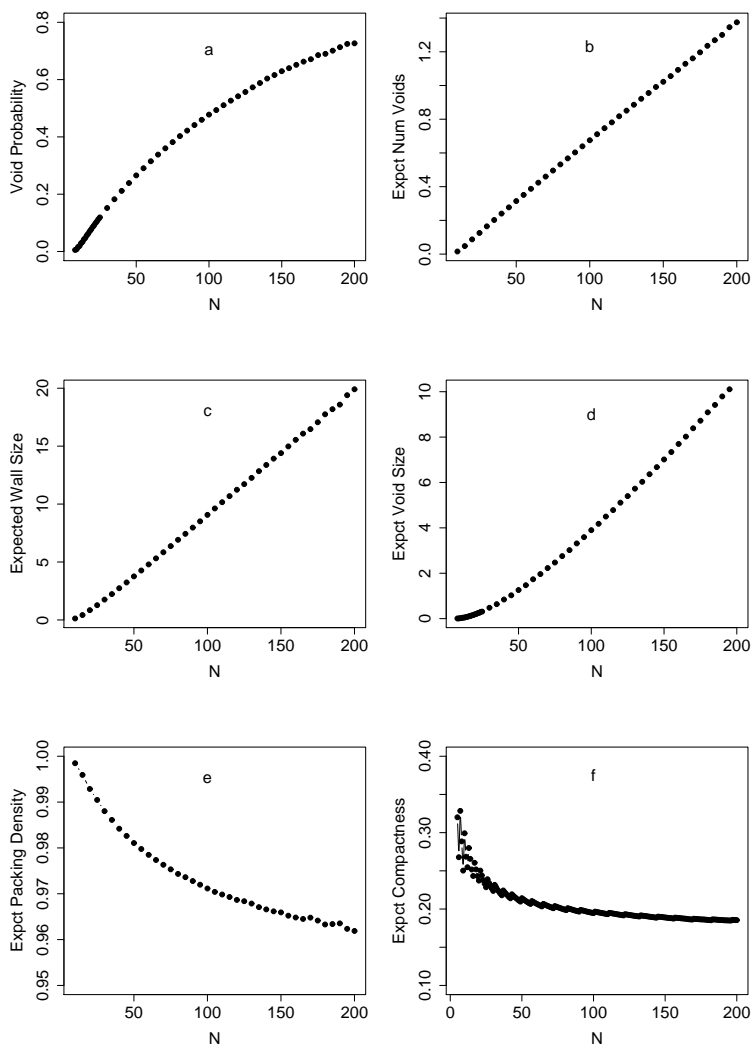,width=4.in}}
\vspace{5mm}
\caption{\sf \footnotesize Geometric properties of lattice polymers of
different lengths, estimated by sequential Monte Carlo method with
2-step look-ahead and resampling technique.  Each Monte Carlo
simulation starts with a sample size of 200,000.  Averaged values of
twenty simulations are shown.  (a) The probability of void formation
increases with chain length. Standard deviation increases slowly with
the length.  At chain length 200, the standard deviation ($8.5 \times
10^-3$) is maximum;  The expected number of voids (standard
deviations $\le 1.6\times10^-2$) (b)  and wall size (standard deviations $\le 0.25$) (c) are linearly correlated with chain length; (d) The
expected void size increases with chain length (standard deviations
$\le 8.3\times10^-3$); (e) The expected packing density decreases with
chain length (standard deviations $\le 7.5\times10^-4$);  (f) The
expected compactness decreases with chain length and reaches an
asymptotic value of $\rho = 0.18$ (standard deviations $\le
5.7\times10^-4$).  Different resampling strategies are applied where
dead conformations are removed and other conformations with the
targeted property is duplicated. Resampling favors conformations with small
radius-of-gyration in (a), (b), (c), (d), (e), and conformations with
large weight in (f).  Resampling is carried out every 5 steps in the
process of chain growth.  }
\label{properties.MC}
\end{figure}

We apply the techniques of sequential Monte Carlo with resampling to
study the statistical geometry of voids in long chain polymers.
Figure~\ref{properties.MC}a shows that the probability of void
formation increases with the chain length.  At chain length 105--110,
about half of the conformations contain voids.  The expected number of
voids (Figure~\ref{properties.MC}b) increases linearly with chain
length.  Similar linear scaling behavior is also observed in proteins
\cite{LiangDill01_BJ}. The expected wall size of void and void size
also increase with chain length (Figure~\ref{properties.MC}c and
Figure~\ref{properties.MC}d).

The expected packing density is found to decrease with chain length,
which is consistent with the scaling relationship of void size and
chain length shown in Figure~\ref{properties.MC}c.  The compactness
$\rho$ of chain polymer has been the subject of several studies
\cite{Ishinabe86_JCP,Chan89_M}.  The asymptotic value of $\rho$ we
found is 0.18, slightly different from that reported in
\cite{Ishinabe86_JCP} ($\rho = 0.16$), and is within the range of 0.16
-- 0.24 reported in \cite{Chan89_M}.

{\bf (Figure~\ref{properties.MC} here.)}

\begin{figure}[t]
  \centerline{\epsfig{figure=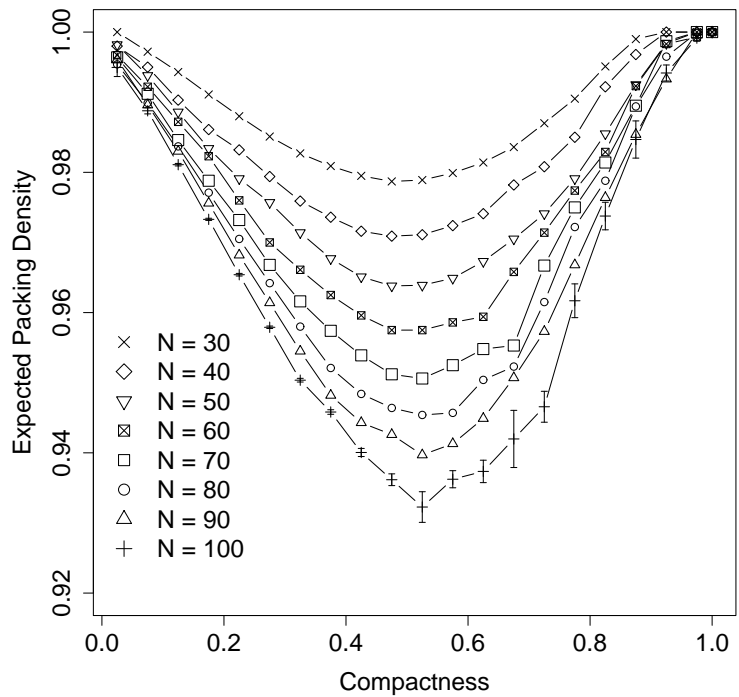,width=4in}}
\vspace{4cm}
\caption{\sf \footnotesize The relationship of expected packing
density and compactness for long chain polymer.  These data are
estimated by sequential Monte Carlo method using 2-step look-ahead and
a sample size of $20\times 200,000$ with resampling.  Resampling is
designed to favor compactness at specified values.  The epxected
packing density calculated by averaging from the 20 runs has the
largest standard deviations for $100$-mer, and are shown in the
figure.}
\label{pd_compactness}
\end{figure}

To explore the relationship of packing density $p$ and compactness
$\rho$, we use sequential Monte Carlo with 2-step look-ahead to sample
200,000 conformations, each with appropriate weight assigned.  This is
repeated 20 times, and the weighted average values of packing density
at various compactness for chains with 60--100 monomers are plotted
(Figure~\ref{pd_compactness}).  The compactness value corresponding to
the minimum packing density seems to have shifted from $0.462$ for
22-mer by enumeration to above 0.5 for 100-mer by sampling.  However,
the overall pattern of $p$ and $\rho$ found by Monte Carlo is very
similar to the pattern found by enumeration for polymers with $N<22$.

The accuracy of geometric properties of long chain polymers estimated
by Monte Carlo can be assessed by the variance obtained from multiple
Monte Carlo runs.

{\bf (Figure~\ref{pd_compactness} here.)}

\section{End Effects of Void Formation}
\begin{figure}[thb]
  \centerline{\epsfig{figure=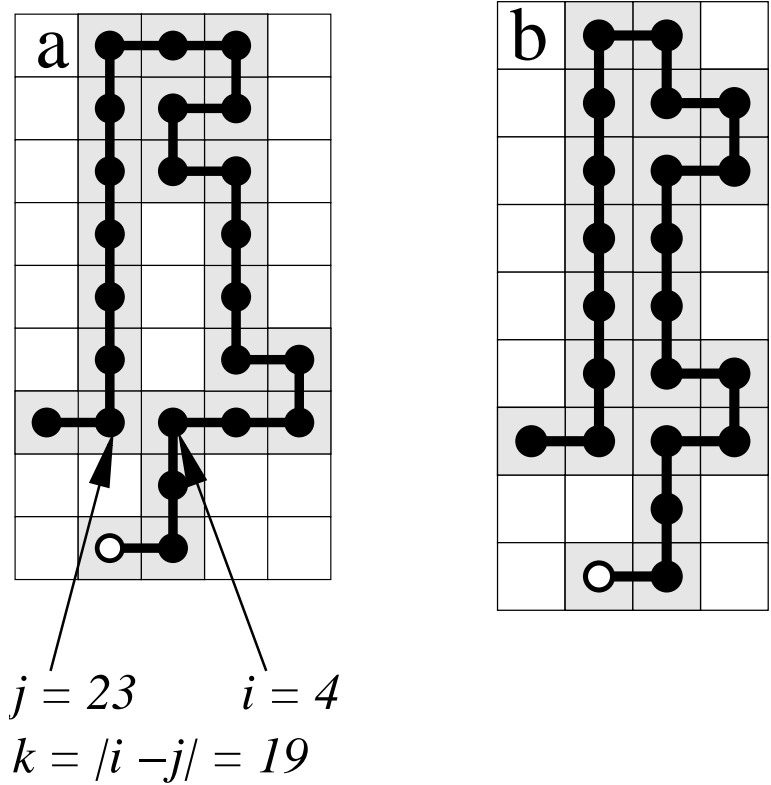,width=4in}}
\vspace{4cm}
\caption{\sf \footnotesize The starting position of a void and its
wall interval.  (a) This 24-mer has a void that starts at $i=4$ and
end at $j=23$.  Its wall size is $k=19$. (b) This polymer has contact-loop but contains no void. }
\label{explain}
\end{figure}

What is the effect of void formation on the size of conformational
space?  We consider the conformational reduction factor of voids.
Following \cite{Chan89_M,ChanDill89_JCP,ChanDill90_JCP}, we define the
conformational reduction factor due to the constraint of a void as:
\[
R(n; i,j) = \frac{\omega(n; i,j)}{\omega(n)},
\]
where $\omega(n; i,j)$ is the number of conformations that contains a
void beginning at monomer $(i)$ and ending at monomer $(j)$, and
$\omega(n)$ is the total number of conformations of
$n$-polymers. $R(n; i,j)$ reflects the restriction of conformational
space due to the formation of a void with wall interval of $k=|i-j|$.
Figure~\ref{explain}a shows a 24-mer with one void that starts at $i =
4$ and $k = 19$.  Unlike self-contacts or self-loops, which was
subject of detailed studies by Chan and Dill
\cite{Chan89_M,ChanDill89_JCP,ChanDill90_JCP}, all conformations
analyzed here must contain a void.  The polymer shown in
Figure~\ref{explain} with a large loop has no void, and such polymers
do not contribute to the numerator of $R$.

{\bf (Figure~\ref{explain} here.)}

\begin{figure}[thb]
  \centerline{\epsfig{figure=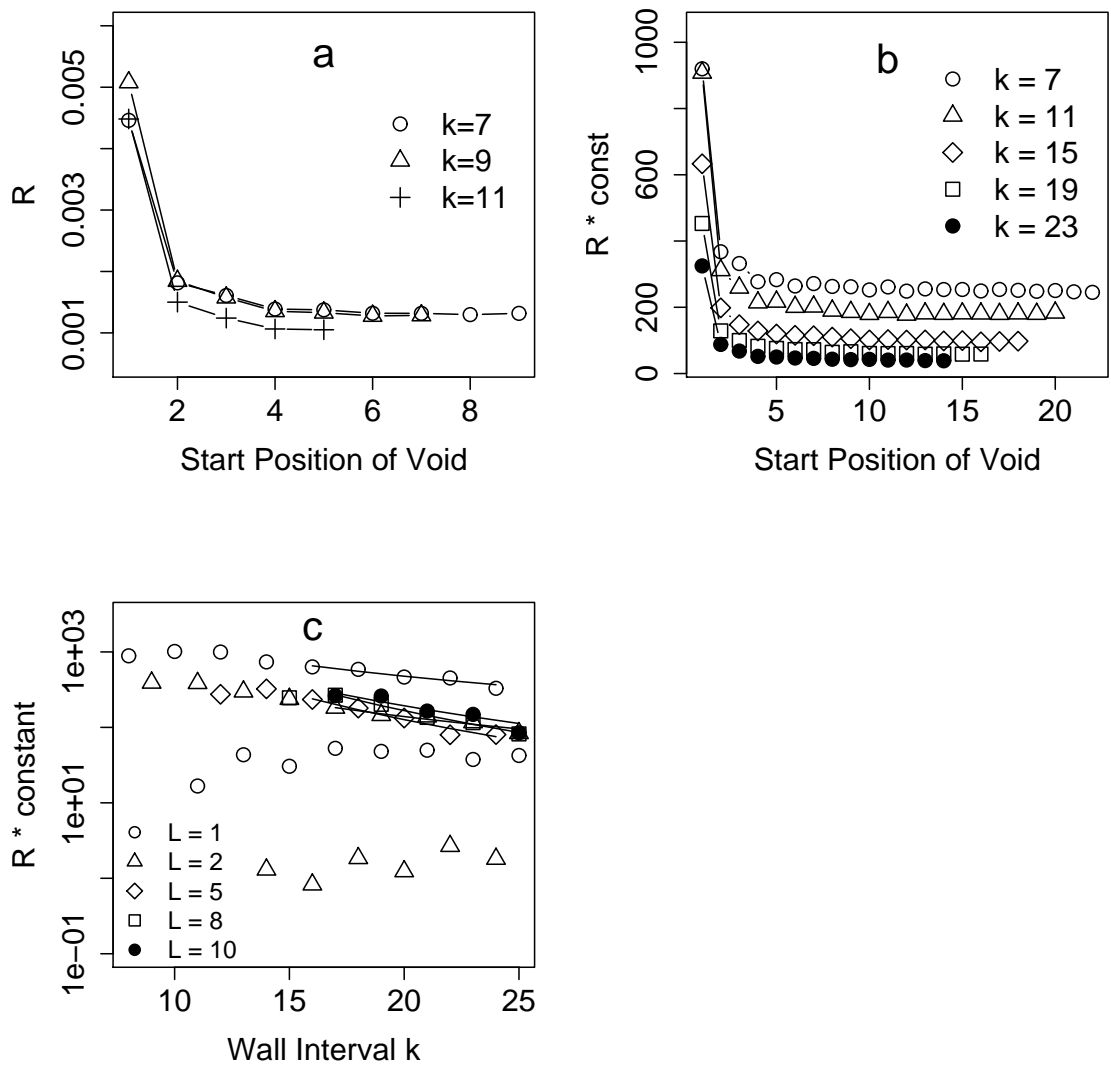,width=6in}}
\vspace{4cm}
\caption{\sf \footnotesize The end-effect of void formation on
conformational reduction. (a) Conformational reduction factor $R$ when
voids are formed in a 22-chain as examined by enumeration.  $R$
depends on the starting position and the wall interval of void; (b)
Conformational reduction factor $R$ upto a normalizing constant when
voids are formed in a 50-chain as sampled by sequential Monte Carlo
(standard deviations $\le 6.2$).  (c) Scaling of conformational
reduction factor $R$ and the wall interval $k$ for 50-chain. (standard deviations $\le 6.4$).
}
\label{EndEffect}
\end{figure}

Figure~\ref{EndEffect}a shows the reduction factor $R$ calculated by
enumeration for voids at different starting positions with wall
intervals $k = 7, 9$ and $11$.  There are clearly strong end-effects:
The reduction factor of voids of the same wall interval depends on
where the void is located.  $R$ decreases rapidly as
the void moves from the end of chain towards the middle.  Void
formation is much more preferred at the end of chain.  Similar end
effects of void formation are also observed for 55-mer sampled by
sequential Monte Carlo (Figure~\ref{EndEffect}b).

{\bf (Figure~\ref{EndEffect} here.)}

The end-effect of voids has the same origin as the end-effect of
self-contact, which has been extensively studied by Chan and Dill
\cite{Chan89_M,ChanDill89_JCP,ChanDill90_JCP}. Because of the effect
of excluded volume, sterically it is less hindering to form a void at
the end of a polymer.  When a void is formed, the conformational space
of the $k+1$ monomers between monomer $i$ and $j$, as well as the two
tails become restricted.  When void is formed at chain end, only one
tail is subject to conformational restriction.

Void formation is different from self-contact.  When monomer $i$ and
$j$ form self-contact, it may involve the formation of a void, but it
is also possible that there will be no unfilled space between $i$ and
$j$.  When a void is formed beginning at monomer $i$ and ending at
monomer $j$, some monomers between $i$ and $j$ will have unsatisfied
contact interactions.  Compare to non-bonded self-contact, the effect
of conformation reduction is more pronounced for void formation.  For
two-dimensional lattice, the ratios between reduction factors of
self-contact at chain end and mid chain of a sufficiently long polymer
are $1.3, 1.4, 1.5$ and $1.6$ for $k = 3, 5, 7$ and $9$, respectively
\cite{ChanDill90_JCP}, whereas the ratios for voids at chain end and
the symmetric midpoint of $N=22$ polymer are $3.4, 4.0$, and $4.4$.
for $k =7, 9$ and $11$.
The conformational reduction factor $R(i,j)$ for voids at various
beginning positions $i$ and various ending position $j$ can be
summarize in a two-dimensional contour plot as shown in
Figure~\ref{EndEffect}b.

We now consider the power-law dependence of $R(N; i, j)$ on the wall
interval $k=|i-j|$.  In the studies of self-contacting loops by Chan
and Dill \cite{ChanDill89_JCP}, the scaling exponent $\nu$ of the
reduction factor $R$ and loop length $k = |i-j|$ for $R(N; i, j)
\approx k^{-\nu}$ is found to be dependent both on $k$ and the
location of the cycle in the chain.  The values of $\nu$ for
self-contact range from 1.6 when $k=N$ to 2.4 when the loop is in the
middle of a long chain with two long tails.  Because void formation
involves at least 8 monomers, its scaling behavior is less amenable to
exhaust enumeration, and application of Monte Carlo sampling is
essential.  Based on estimations from Monte Carlo simulation of void
formation in 50-mer, $\nu$ ranges from $1.4\pm0.2$ for $l_0 =1$ to
$3.0\pm0.2$ for void initiation position $l_0 = 8$
(Figure~\ref{EndEffect}c).  Our results show that the scaling exponent
of $R$ with $k=|i-j|$ for void formation is similar to that of
self-contacting loop.  This scaling exponent also depends on the
location of the void.

\section{Conclusion}
In this work, we have studied the statistical geometry of voids as
topological features in two-dimensional lattice chain polymers.  We
define voids as unfilled space fully contained within the polymer, and
have developed a simple algorithm for its detection.  We have explored
the relationship of various statistical geometric properties with the
chain length of the polymer, including the probability of void
formation $\pi_v$, the expected number of voids $\bar{n}_v$, the
expected void size $\bar{v}$, the expected wall size of voids
$\bar{w}$, packing density $p$, and the expected compactness $\rho$.
Our results show that for chains of $>$105-110 monomers, at least half
of the conformations contain a void.  At about 150 monomers, there
will be at least one void expected in a polymer.  The expected wall
size scale linearly with the chain length, and about 10\% of the
monomers participate in the formation of voids.  We formalize the
concept of packing density for lattice polymers.  We found that both
the packing density and compactness decrease with chain length.  The
asymptotic value of compactness $\rho$ is estimated to be 0.18.

We have also characterized the relationship of packing density and
compactness, two parameters that have been used frequently for
studying protein packing.  Our results indicate that packing density
reaches minimum values between compactness 0.4 -- 0.6.  The entropic
effects of voids are studied by analyzing the conformational reduction
factor $R$ of void formation. We found that there is significant
end-effect for void formation: the ratio of $R$ at chain end and at
mid chain may be twice as large as that of the $R$ factor for contact
loops, where the formation of voids is not required.

\begin{figure}[t]
  \centerline{\epsfig{figure=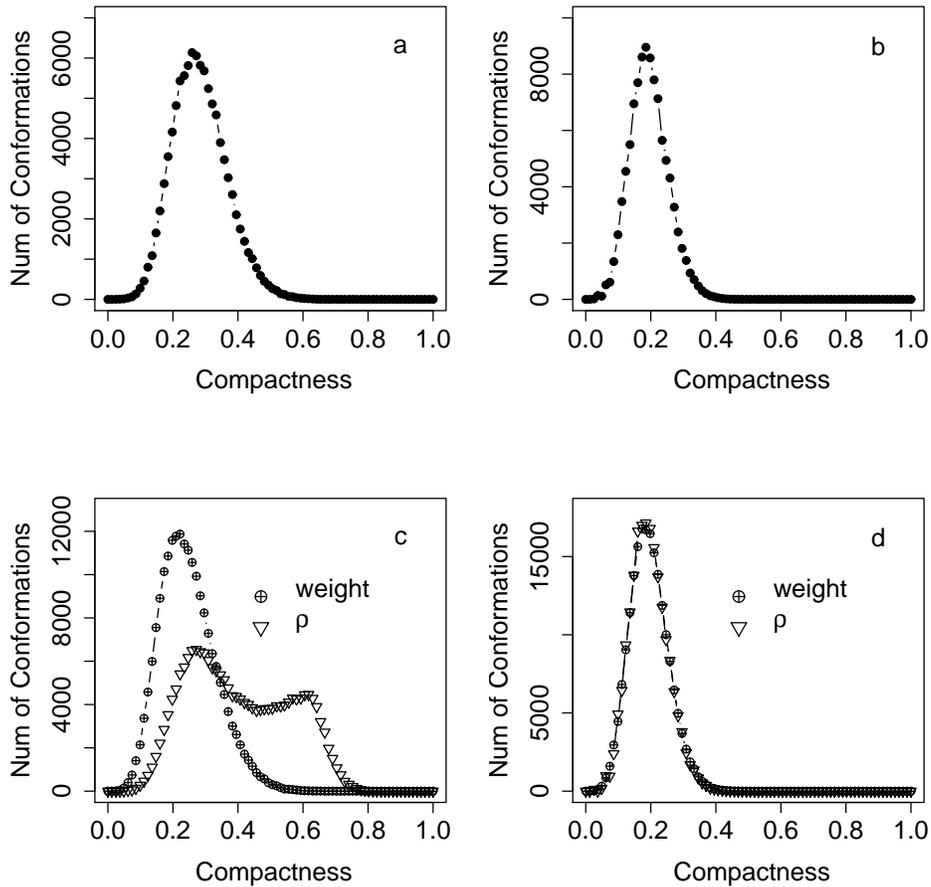,width=5.in}}
\vspace*{8mm}
\caption{\sf \footnotesize Histograms of conformations of 100-mers
generated by sequential Monte Carlo with and without resampling at
different compactness.  In (c) and (d), resampling is applied to every
step of the chain growth process.  All weighted histogram is
normalized so the total area equals to the total number of surviving
conformations reaching 100-mer.  (a) Histogram of conformations at
different compactness generated without resampling.  (b) Weighted
histogram of conformations generated without resampling, which is
proportional to the distribution of all geometrically feasible
100-mers.  (c) Histograms of conformations at different compactness
when resampling is applied.  To resample by weight, dead conformations
are replaced with conformations of highest weight.  To resample by
compactness, dead conformations are replaced with conformations of
lowest compactness.   Note that the total number
of surviving conformations that reach chain length of 100 is much
higher then without resampling.  Resampling by compactness generates
many more conformations with higher compactness.  (d) The weighted
histograms of conformations under different resampling are in
excellent agreement with each other and with that when no resampling
is applied.    }
\label{weightHist}
\end{figure}

In this study, we have applied sequential Monte Carlo sampling and
resampling techniques to study the statistical geometry of voids.
Sequential Monte Carlo sampling and resampling is essential for
exploring the geometry of long chain polymers.  This is a very general
approach that allows the generation of increased number of
conformations with interesting characteristics.  For example, we can
replace dead conformations with existing conformations of highest
weight, or conformations with highest compactness, or with smallest
radius of gyration.  Figure~\ref{weightHist}a shows the histograms of
conformation of 100-mer at different packing density generated without
resampling.  Figure~\ref{weightHist}c shows the histograms of
conformations when resampling by weight and resampling by compactness
$\rho$ are used.  Other resampling schemes are possible, {\it e.g.}
resampling by radius-of-gyration, by packing density. During
resampling, the number $k$ of dead conformations at each step of
growth is identified and these are replaced with conformations of
interest from $k$ randomly divided groups.  These conformations must
have not been resampled in previous 4 steps of the growth process to
maintain sample diversity.  Both histograms where resampling is used
deviate from that of Figure~\ref{weightHist}a.  Resampling by weight
shifts the peak of the conformations to below 0.2, and resampling by
compactness turns the histogram into bi-modal.  The latter produces a
lot more conformations with compactness $\rho>0.4$.

SMC sampling and resampling use biased samples since conformations are
generated with probability different from that of the target
distribution.  The bias is ictated by different method of resampling
and different choices of the number of steps of look-ahead in
sequential Monte Carlo.  An essential component of a successful biased
Monte Carlo sampling is the appropriate weight assignment to each
sample conformation.  This is necessary because we need to estimate
the expected values of parameter such as packing density and void size
under the target distribution of all geometrically feasible
conformations.  In Figure~\ref{weightHist}a where each of the 200,000
starting conformations is
generated by two-step look-ahead without resampling, not every
conformation is generated with the same probability and therefore is
assigned different weight accordingly.  Figure~\ref{weightHist}b shows
the weight-adjusted histogram, which is indicative of the probability
density function at different compactness for the population of all
geometrically feasible 100-mers.

Figure~\ref{weightHist}d shows that when weights are incorporated and
the area of the histogram normalized to the final number of surviving
conformations, the weighted distributions of conformations using
different resampling techniques have excellent agreement with the
weighted distribution when no resampling is used
(Figure~\ref{weightHist}b).  This example shows that by incorporating
weights, the target distributions can be faithfully recovered even
when the sampling is very biased.

{\bf (Figure~\ref{weightHist} here.)}

Although sequential Monte Carlo sampling is very effective, the
estimation of parameters associated with rare events remain difficult.
In Figure~\ref{EndEffect} where conformational reduction factor $R$ is
plotted at various void initiation position and wall interval length,
voids starting at position of 1 but with odd wall intervals ($k \in
\{11, 13, ..., 25\}$) are much rarer, and it is unlikely sequential
Monte Carlo sampling with limited sample size can provide large enough
effective sample size for the accurate estimation of scaling
parameters $\nu$, where $R(N; i,j) \approx k^{-\nu}$.

In this study, we are interested in the statistics of void geometry,
and our target distribution is the uniform distribution of all
conformations of length $n$.  With the introduction of appropriate
potential function and alphabet of monomers such as the HP model
\cite{Dill85_Biochem,LauDill89_M,ChanDill93_JCP}, we can study the
thermodynamics, kinetics, and sequence degeneracy of chain polymers
when voids are formed in polymers. In these cases, our target
distributions will be chain polymers under the Boltzmann distribution
derived from the corresponding potential functions.

\section{Acknowledgments}
This work is supported by funding from National Science Foundation
DMS 9982846, CMS 9980599, DMS 0073601, DBI0078270, and MCB998008,
and American Chemical Society/Petroleum Research Fund.

\newpage
\section{Appendix}
To detect voids in a polymer, we use a simple search method.  For an  $l
\times l$ lattice, we start from the lower-left corner.  Once we found
an unoccupied site $u$, we use the breadth-first-search (BFS) method
to identify all other unoccupied sites that are connected to site
$u$. These sites are grouped together and marked as ``visited''.
Collectively they represent one void in the lattice.  We continue this
process until all unoccupied sites are marked as visited:

\begin{tabbing}
123\=456\=789\=\kill
{\bf Algorithm} {\sc VoidDetection} ({\it lattice}, $l$)\\
$v=0$  {\sf // Number of voids}\\
{\tt for } $i =1$ {\tt to} $l$ \\
   \>{\tt for } $j =1$ {\tt to} $l$ \\
   \>   \> {\tt if} site($i,j$) is unoccupied and not visited\\
   \>   \>   \>$v \leftarrow v+1$\\
   \>   \>   \>Mark ($i, j$) as visited.\\
   \>   \>   \>{\sc BreadthFirstSearch}({\it lattice}, ($i, j$))\\
   \>   \>   \>Update the size of  $void(i, j)$\\
   \>   \> {\tt endif}\\
   \>{\tt endfor}\\
{\tt endfor}
\end{tabbing}
Details of BFS can be found in algorithm textbooks such as
\cite{Cormen90}. 

\newpage
\bibliography{svm,array,prop,pack,prf,lattice,bioshape,liang,potential,bix,more,express,whit}
\bibliographystyle{unsrt}

\newpage
\begin{table*}[thb] % one column table
 \begin{center}
\caption{
Number of conformations of a $n$-polymer with different number of voids on a square lattice.
}
\label{exhaust.tab}
\vspace*{.9in}
 \begin{tabular}{rrrrrrr}
   $n$ & $\omega(n)$ & $\omega_0(n)$ & $\omega_1(n)$ & $\omega_2(n)$ &
         $\omega_3(n)$ & $\omega_4(n)$    \\ \hline
3   &    2           &      2           &    0          &     0      & 0      & 0   \\
4   &    5           &      5           &    0          &     0      & 0      & 0   \\
5   &    13          &      13          &    0          &     0      & 0      & 0   \\
6   &    36          &      36          &    0          &     0      & 0      & 0   \\
7   &    98          &      98          &    0          &     0      & 0      & 0   \\
8   &    272         &      270         &    2          &     0      & 0      & 0   \\
9   &    740         &      734         &    6          &     0      & 0      & 0   \\
10  &    2034        &      1993        &    41         &     0      & 0      & 0   \\
11  &    5513        &      5393        &    120        &     0      & 0      & 0   \\
12  &    15037       &      14508       &    529        &     0      & 0      & 0   \\
13  &    40617       &      39078       &    1536       &     3      & 0      & 0   \\
14  &    110188      &      104566      &    5602       &     20     & 0      & 0   \\
15  &    296806      &      280599      &    16088      &     119    & 0      & 0   \\
16  &    802075      &      748335      &    53149      &     591    & 0      & 0   \\
17  &    2155667     &      2002262     &    151052     &     2353   & 0      & 0   \\
18  &    5808335     &      5327888     &    470386     &     10051  & 10     & 0   \\
19  &    15582342    &      14222389    &    1325590    &     34287  & 76     & 0   \\
20  &    41889578    &      37784447    &    3973361    &     131298 & 472    & 0   \\
21  &    112212146   &      100673771   &    11119456   &     416239 & 2680   & 0   \\
22  &    301100754   &      267136710   &    32479871   &     1471874& 12293  & 6   \\
23  &    805570061   &      710673806   &    90361878   &     4479355& 54998  & 24  \\
24  &    2158326727  &      1883960171  &    259195774  &     14946910 & 223458 & 414  \\
25  &    5768299665  &      5005591512  &    717505892  &     44337381 & 862748     & 2132  \\
 \end{tabular}
 \end{center}
\end{table*}

\newpage

\newpage
\begin{figure}
\vspace{2cm}
  \centerline{\epsfig{figure=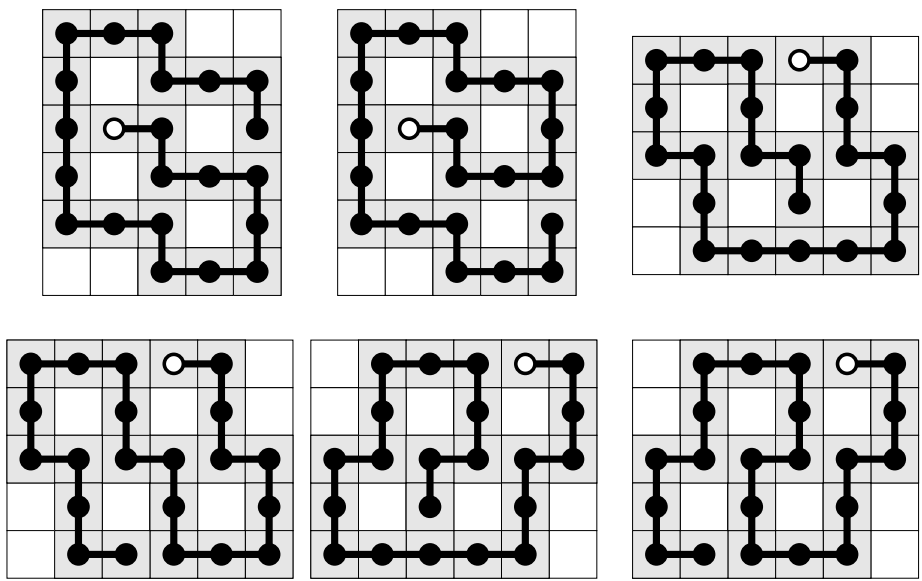,width=5in}}
\vspace{4cm}
\caption{\sf \footnotesize The only six conformations of $22$-mer that
contain $4$ voids.  }
\label{22merVoids4}
\end{figure}

\newpage

\newpage

\newpage

\newpage

\newpage

\newpage

\newpage

\newpage

\newpage

\newpage

\end{document}